\documentclass[twocolumn,showpacs,amsmath,amssymb,superscriptaddress,pra]{revtex4}
\usepackage{graphicx}
\usepackage{dcolumn}
\usepackage{mathrsfs}
\usepackage{bm,amsmath}
\usepackage{times}

\begin{document}

\title{Proposal for implementing universal superadiabatic geometric quantum gates in nitrogen-vacancy centers }

\author{Zhen-Tao Liang}
\affiliation{Guangdong Provincial Key Laboratory of Quantum
Engineering and Quantum Materials, School of Physics and
Telecommunication Engineering, South China Normal University,
Guangzhou 510006, China}

\author{Xianxian Yue}
\affiliation{Guangdong Provincial Key Laboratory of Quantum
Engineering and Quantum Materials, School of Physics and
Telecommunication Engineering, South China Normal University,
Guangzhou 510006, China}

\author{Qingxian Lv}
\affiliation{Guangdong Provincial Key Laboratory of Quantum
Engineering and Quantum Materials, School of Physics and
Telecommunication Engineering, South China Normal University,
Guangzhou 510006, China}

\author{Yan-Xiong Du}
\affiliation{Guangdong Provincial Key Laboratory of Quantum
Engineering and Quantum Materials, School of Physics and
Telecommunication Engineering, South China Normal University,
Guangzhou 510006, China}

\author{Wei Huang}
\affiliation{Guangdong Provincial Key Laboratory of Quantum
Engineering and Quantum Materials, School of Physics and
Telecommunication Engineering, South China Normal University,
Guangzhou 510006, China}

\author{Hui Yan}
\email{yanhui@scnu.edu.cn} \affiliation{Guangdong Provincial Key
Laboratory of Quantum Engineering and Quantum Materials, School of
Physics and Telecommunication Engineering, South China Normal
University, Guangzhou 510006, China}

\author{Shi-Liang Zhu}
\email{slzhu@nju.edu.cn}
 \affiliation{National Laboratory of
Solid State Microstructures, School of Physics, Nanjing University,
Nanjing 210093, China}

 \affiliation{Guangdong Provincial Key
Laboratory of Quantum Engineering and Quantum Materials, School of
Physics and Telecommunication Engineering, South China Normal
University, Guangzhou 510006, China}

\affiliation{Synergetic Innovation Center of Quantum Information and
Quantum Physics, University of Science and Technology of China,
Hefei, Anhui 230026, China}


\begin{abstract}
We propose a feasible scheme to implement a universal set of quantum
gates based on  geometric phases and superadiabatic quantum control.
Consolidating the advantages of both strategies, the proposed
quantum gates are robust and fast. The diamond nitrogen-vacancy
center system is adopted as a typical example to illustrate the
scheme. We show that these gates can be realized in a simple
two-level configuration by appropriately controlling the amplitude,
phase, and frequency of just one microwave field. The gate's robust
and fast features are confirmed by comparing the fidelity of the
proposed superadiabatic geometric phase (controlled-PHASE) gate with
those of two other kinds of phase (controlled-PHASE) gates.
\end{abstract}


\maketitle

Because the diamond nitrogen-vacancy (NV) center system has the
potential to operate at room temperature, it has attracted a lot of
interest in quantum computation
\cite{Wrachtrup,Neumann,Yao,Dolde1,Dolde2} and quantum sensing
\cite{Balasubramanian,Maze,Taylor,Dolde} research in recent years.
In order to realize scalable quantum computation based on NV center
systems, the fidelity of each quantum gate needs to exceed a certain
threshold \cite{Knill,Dawson}. Geometric phases are believed to be
robust against local stochastic noises, which
 depend solely on certain global geometric features of the executed
evolution paths \cite{Chiara,Zhu1,Berger}. Therefore, quantum gates
and quantum control based on geometric phases, which are
intrinsically fault tolerant \cite{E,Tan}, have been under
consideration for the NV center systems. Very recently, geometric
quantum gates based on purely nonadiabatic geometric phases were
realized in room temperature NV center systems
\cite{Arroyo-Camejo,Zu} as well as other systems
\cite{Leibfried,Abdumalikov}. Compared with the initial schemes for
geometric quantum computation \cite{Zanardi} based on adiabatic
non-Abelian holonomies \cite{Duan}, the nonadiabatic geometric
phases \cite{Zhu2,Wang} and nonadiabatic non-Abelian holonomies
\cite{ES,Liang,Zhou,Xue,Xu1,Xu2} allow for high-speed quantum gate
operations and thus intrinsically protect against
environment-induced decoherence such as decay and dephasing.
However, the nonadiabatic geometric quantum gates are susceptible to
the systematic errors in the control Hamiltonian
\cite{Thomas,Johansson,Wu}.

On the other hand, the superadiabatic and counteradiabatic
\cite{Lim,Berry,Demirplak1,Demirplak2,Demirplak3,Santos,Zhang2}
quantum control proposals are believed to be not only remarkably
fast (with speed close to the quantum speed limit) but also highly
robust against control parameter variations. In the superadiabatic
scheme, the system evolves exactly along the instantaneous
eigenstate of an original Hamiltonian at any desired rate by
introducing an additional Hamiltonian \cite{Chen,Giannelli,Campo}.
Interestingly, high-fidelity, robust and fast quantum control based
on the superadiabatic protocol has been experimentally realized on
the cold atomic ensemble \cite{Du1}, the NV electron qubit
\cite{Zhang}, and the atomic optical lattice system \cite{Bason}. It
is noteworthy that as adiabatic population transfers
\cite{Bergmann,Klein,Du2}, superadiabatic population transfers are
insensitive to the dynamical evolution times, so it is not even
necessary to design  the exact durations of the controlling fields
beforehand.

For scalable quantum computation, the ideal quantum gates should
be both robust and fast.
Since geometric quantum manipulation has an intrinsical fault
tolerant property \cite{Berger,Chiara,Zhu1,E,Tan} and superadiabatic
control is remarkably fast \cite{Du1,Zhang,Bason}, the realization
of a universal set of quantum gates which consolidates the
aforementioned advantages of geometric phases and superadiabatic
evolution will be essential in quantum computation and quantum
manipulation.

In this paper, we propose an experimentally feasible scheme to
implement universal superadiabatic geometric quantum gates (SGQGs)
that are both robust and fast. The scheme can be used in many
candidates of quantum computation. For the purpose of demonstration,
we adopt the NV center system as a typical example to illustrate
this approach. This system is controlled by microwave fields. We
show that a universal set of SGQGs can be realized in a simple
two-level configuration by appropriately controlling the amplitude,
phase, and frequency of just one microwave field.
On one hand, the evolutions are geometric and thus robust against
 certain high-frequency fluctuations. On the other hand, the
evolutions are superadiabatic and can thus be fast and robust
against systematic errors. We  compare the fidelities of the
proposed SGQGs with those of two other kinds of phase gates, and
find that the SGQGs can perform ten times faster but with the
fidelities comparable with that of the normal adiabatic geometric
phase gate. The fidelities and operation time are comparable with
those of the nonadiabatic holonomic gates which are implemented in a
three-level structure.

We first address a general approach to achieve SGQGs using a
two-level energy structure. We assume that the two-level system
couples with a microwave field with frequency $\omega_{m}$ and
phase $\varphi$. The energy difference between the states
$|0\rangle$ and $|1\rangle$ is $\hbar\omega$. The system is
described by a time-dependent Hamiltonian
\begin{equation}
H_{0}(t)=\frac{\hbar}{2}\left(\begin{array}{ccc}
\Delta(t)&\Omega_{R}(t)e^{-i\varphi(t)}\\
\Omega_{R}(t)e^{i\varphi(t)}&-\Delta(t)
\end{array}\right),
\end{equation}
where $\Delta(t)=\omega-\omega_{m}(t)$ is the detuning and
$\Omega_{R}(t)$ (real here and hereafter) is the Rabi frequency
proportional to the amplitude of the microwave field. The
instantaneous eigenstates are
$$|\lambda_{+}(t)\rangle=\left( \begin{array}{c} \cos \frac{\theta}{2}e^{-i\varphi/2} \\ \sin \frac{\theta}{2} e^{i\varphi/2} \end{array} \right),
\ \ \
|\lambda_{-}(t)\rangle=\left( \begin{array}{c}
-\sin \frac{\theta}{2} e^{-i\varphi/2} \\
\cos \frac{\theta}{2}e^{i\varphi/2} \end{array} \right),
$$ where the mixing angle
$\theta=\arctan[\Omega_{R}(t)/\Delta(t)]$. The corresponding
eigenvalues are $E_{\pm}=\pm\hbar\Omega/2$, where
$\Omega=\sqrt{\Delta^{2}(t)+\Omega^{2}_{R}(t)}$. With adiabatic
approximation, the time evolution takes the following form:
$U(t)=\sum_n
\exp\{i\int^{t}_{0}[A_{n}(t^{\prime})-E_{n}(t^{\prime})/\hbar]dt^{\prime}
\}|\lambda_{n}(t)\rangle\langle\lambda_{n}(0)|$ $(n=+,-)$, where
$A_{n}(t^{\prime})=i\langle\lambda_{n}(t^{\prime})|\partial_{t^{\prime}}\lambda_{n}(t^{\prime})\rangle$
is the effective vector potential \cite{Berry2}. Through the
reverse engineering approach \cite{Berry}, a single superadiabatic
Hamiltonian $H_{s}(t)=H_{0}(t)+H_{1}(t)$, where
$$H_{1}(t)=i\hbar\sum\limits_{n=+,-}[|\partial_{t}\lambda_{n}\rangle\langle
\lambda_{n}|-\langle
\lambda_{n}|\partial_{t}\lambda_{n}\rangle|\lambda_{n}\rangle\langle
\lambda_{n}|]$$
 is the superadiabatic correction Hamiltonian, will
drive all the eigenstates $|\lambda_{n}(t)\rangle$ of $H_{0}(t)$
to precisely evolve without transitions between them at any
desired rate.

We can further deduce the superadiabatic Hamiltonian. For the
given system driven by Hamiltonian (1), when $\varphi$ is kept
constant, the superadiabatic correction Hamiltonian reads
\begin{equation}
H_{1}(t)=\frac{\hbar}{2}\left(\begin{array}{lll}
0&-i\Omega_{c}(t)e^{-i\varphi}\\
i\Omega_{c}(t)e^{i\varphi}&0
\end{array}\right),
\end{equation}
where
$\Omega_{c}(t)\equiv\dot{\theta}=[\dot{\Omega}_{R}(t)\Delta(t)-\Omega_{R}(t)\dot{\Delta}(t)]/\Omega^{2}$.
In order to ensure that the system evolves perfectly adiabatically,
it seems that another microwave field with phase $\varphi+\pi/2$ is
required \cite{Chen}, which may increase the difficulty of
experiments. However, the effect of this extra microwave field can
be achieved by appropriately modifying the Rabi frequency
(amplitude) and the phase of the original microwave field. 
Therefore, we recast the wanted Hamiltonian
$H_{s}(t)=H_{0}(t)+H_{1}(t)$ in the form
\begin{equation}
\label{Hs} H_{s}(t)=\frac{\hbar}{2}\left(\begin{array}{ccc}
\Delta(t)&\Omega_{s}(t)e^{-i[\varphi+\phi_{s}(t)]}\\
\Omega_{s}(t)e^{i[\varphi+\phi_{s}(t)]}&-\Delta(t)
\end{array}\right),
\end{equation}
where $\Omega_{s}(t)=\sqrt{\Omega^{2}_{R}(t)+\Omega^{2}_{c}(t)}$ and
$\phi_{s}(t)=\arctan[\Omega_{c}(t)/\Omega_{R}(t)]$, which eliminates
the need for an extra microwave field to realize the $H_{1}(t)$
term.

The states $|\lambda_\pm (t) \rangle$ are a pair of orthogonal
states and can be used to realize SGQGs under the cyclic condition
$|\lambda_\pm (T) \rangle=e^{i\phi_\pm}|\lambda_\pm (0) \rangle$,
where $\phi_\pm$ are real phase factors. We first denote an
arbitrary initial state as $|\psi_{i}\rangle=a_{+}|\lambda_{+} (0)
\rangle+a_{-}|\lambda_{-} (0)\rangle$ with
$a_{\pm}=\langle\lambda_{\pm} (0)|\psi_{i}\rangle$. Then, we
cyclically change the superadiabatic Hamiltonian $H_{s}(t)$ with
period $T$ by suitably manipulating the parameters $\Delta(t)$,
$\Omega_R (t)$, and $\varphi (t)$. States $|\lambda_{+}\rangle$ and
$|\lambda_{-}\rangle$ thus evolve cyclically and gain different
phases including both geometric and dynamical components. If a
special cyclic evolution path is chosen to erase the accumulated
dynamical phases, pure geometric phases can be obtained. Under those
conditions, we find the relation
$U(T)|\lambda_{\pm}(0)\rangle=\exp(\pm i\gamma)|\lambda_{\pm} (0)
\rangle$, where $\gamma$ is a pure  geometric  phase given by the
vector potential integral in $U(t)$. The final state at time $T$ is
found to be $|\psi_{f}\rangle=U(\chi,\gamma)|\psi_{i}\rangle$, where
\begin{equation}
\label{SU} U=\left(\begin{array}{ccc}
\cos\gamma+i\cos\chi \sin\gamma& i\sin\chi\sin\gamma  \\
i\sin\chi\sin\gamma & \cos\gamma-i\cos\chi \sin\gamma
\end{array}\right)
\end{equation}
is a single-qubit gate depending only on the geometric phase
$\gamma$ and the initial value $\chi=\theta (0)$ \cite{Zhu2}.
 We note that Eq.
(\ref{SU}) is phase gate $U_1=\exp (i\gamma\sigma_z)$ when
$\chi=0$
 and is $U_2=\exp (i\gamma\sigma_x)$ when $\chi=\pi/2$; any single-qubit operation can be realized by a
combination of $U_1$ and $U_2$. Moreover, a nontrivial conditional
two-qubit gate can also be realized if there exist two different
pairs of orthogonal cyclic states of the target qubit, conditional
upon the state of the control qubit. Since $\gamma$ in Eq.
(\ref{SU}) is a geometric phase realized with a superadiabatic
Hamiltonian, we call it a SGQG. Therefore, we have proposed a
general scheme  to achieve a universal set of SGQGs.


The scheme can be valid in many systems. It is important
to implement two kinds of noncommutable single-qubit gates and one
nontrivial two-qubit gate with realistic physical systems. For
demonstration, we illustrate such an implementation based on the
control of electron and nuclear spins in a diamond NV center with a
proximal $^{13}$C atom. The NV center has a spin-triplet ground
state and the nearby nuclear spins ($^{13}$C and the host $^{15}$N)
are polarized by a magnetic field of about 500 G along the
nitrogen-vacancy axis \cite{Jacques}. Owing to the large energy
difference of $|m_{s}=\pm1\rangle$ levels shifted by the magnetic
field, we take the Zeeman components
$|m_{s}=0\rangle\equiv|0\rangle$ and
$|m_{s}=-1\rangle\equiv|1\rangle$ as the qubit basis states, as
shown in Fig. 1(a). The qubit can be manipulated by a microwave
field whose frequency, amplitude, and phase can be adjusted by
mixing with an arbitrary-waveform generator. Therefore, the
superadiabatic Hamiltonian (\ref{Hs}) can be straightforwardly
realized by driving the microwave field.

\begin{figure}[ptb]
\begin{center}
\includegraphics[width=9cm]{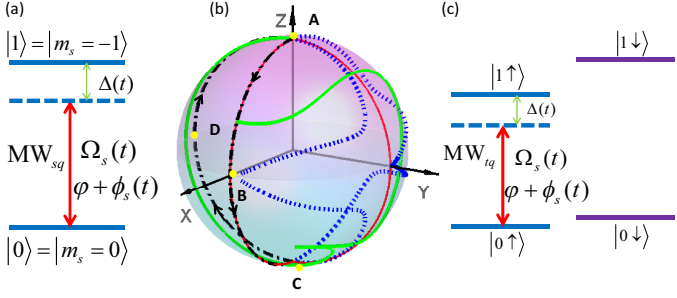}
\caption{\label{fig:charateristic}(color online). Level structure
and ``orange slice" scheme for SGQGs. (a) Two Zeeman levels
$|m_{s}=0\rangle$ and $|m_{s}=-1\rangle$ of the NV spin-triplet
ground state are encoded as the qubit states $|0\rangle$ and
$|1\rangle$. The double-sided red arrow indicates level-selective
coupling of microwave fields for geometric manipulation of the
qubit. (b) Evolution of the cyclic state $|\lambda_{+}\rangle$
starting from point $A$ and driven by the superadiabatic (original)
Hamiltonian is represented by the polarized vector $\mathbf{n}_{+}$
[dot-dashed black line (solid green line)] and the corresponding
effective magnetic field $\mathbf{B}_{s} (\mathbf{B}_{0})$ [dotted
blue line (solid red line)]. (c) The level structure describes the
hyperfine interaction of the NV electron spin with the $^{13}$C
nuclear spin ($I=1/2$). The thick double-sided red arrow indicates
state-selective coupling of microwave fields for the nontrivial
two-qubit gates acting on the computational space spanned by
$\{|0,\downarrow\rangle$, $|1,\downarrow\rangle$,
$|0,\uparrow\rangle$, $|1,\uparrow\rangle\}$.}
\end{center}
\end{figure}

We now show schematically how to construct the phase gate $U_{1}$
in two steps via an ``orange slice" scheme as shown in Fig. 1(b),
where we plot a designed cyclic evolution path
($ABCDA$) on the Bloch sphere surface. 
At this stage, we choose point $A$ (corresponding to $\chi=0$) in
Fig. 1(b) to be the initial state $|\lambda_{+}\rangle$. In order to
drive the state $|\lambda_{+}\rangle$ ($|\lambda_{-}\rangle$) to
evolve cyclically, with the accumulated dynamical phase being zero,
we control the microwave field as follows. In the first step, from
$t=0$ to $t=2\tau$, the Rabi frequency and detuning of the original
Hamiltonian $H_{0}(0,2\tau)$ with a constant phase $\varphi_{1}$ are
given by
\begin{equation}
\Omega_{R}(t)=
\begin{cases}
\Omega_{0}[1-\cos(\frac{\pi t}{\tau})], &\mbox{$0\leq t<\tau$}\\
\Omega_{0}[1+\cos(\frac{\pi (t-\tau)}{\tau})], &\mbox{$\tau\leq,
t\leq2\tau$}
\end{cases}
\end{equation}
and
\begin{equation}
\Delta(t)=
\begin{cases}
\Delta_{0}[\cos(\frac{\pi t}{\tau})+1], &\mbox{$0\leq t<\tau$}\\
\Delta_{0}[\cos(\frac{\pi (t-\tau)}{\tau})-1], &\mbox{$\tau\leq,
t\leq2\tau$}.
\end{cases}
\end{equation}
The system is driven to point $C$ by superadiabatic Hamiltonian
$H_{s}(0,2\tau)$ with a constant phase $\varphi_{1}$. In the second
step, from $t=2\tau$ to $t=4\tau$, the parameters of the Hamiltonian
$H_{0}(2\tau,4\tau)$ with a constant phase $\varphi_{2}$ are given
by
\begin{equation}
\Omega_{R}(t)=
\begin{cases}
\Omega_{0}[1-\cos(\frac{\pi (t-2\tau)}{\tau})], &\mbox{$2\tau\leq t<3\tau$}\\
\Omega_{0}[1+\cos(\frac{\pi (t-3\tau)}{\tau})], &\mbox{$3\tau\leq,
t\leq4\tau$}
\end{cases}
\end{equation}
and
\begin{equation}
\Delta(t)=
\begin{cases}
\Delta_{0}[\cos(\frac{\pi (t-2\tau)}{\tau})+1], &\mbox{$2\tau\leq t<3\tau$}\\
\Delta_{0}[\cos(\frac{\pi (t-3\tau)}{\tau})-1], &\mbox{$3\tau\leq,
t\leq4\tau$}.
\end{cases}
\end{equation}
The system is driven back to point $A$ by superadiabatic Hamiltonian
$H_{s}(2\tau,4\tau)$ with a constant phase $\varphi_{2}$. A crucial
requirement for the geometric quantum  gate is that the dynamical
phase should be vanished. To this end, we denote the polarized
vector
$\mathbf{n}_{\pm}(t)=\langle\lambda_{\pm}(t)|\vec{\sigma}|\lambda_{\pm}(t)\rangle$,
and it is straightforward to find that
$\mathbf{n}_{+}(t)=-\mathbf{n}_{-}(t)$. As shown in Fig. 1(b), the
vector $\mathbf{n}_{+}(t)$ (dot-dashed black line), which is driven
by superadiabatic Hamiltonian $H_{s}$,  follows precisely the
effective magnetic field $\mathbf{B}_{0}(t)$ (solid red line)
yielded from the original Hamiltonian $H_{0}$. Since
$\cos[\mathbf{n}_{+}(t),\mathbf{B}_{s}(t)]=\cos[-\mathbf{n}_{+}(t+2\tau),\mathbf{B}_{s}(t+2\tau)]=\arccos[\sin^{2}\theta(t)\cos\phi_{s}(t)+\cos^{2}\theta(t)]$,
where $\mathbf{B}_{s}(t)$ (dotted blue line) is the effective
magnetic field yielded from the superadiabatic Hamiltonian $H_{s}$,
the dynamic phases accumulated in the paths $ABC$ and $CDA$ are
completely canceled out. Meanwhile, the wanted geometric phase
$\gamma_{1}=\pi-(\varphi_{2}-\varphi_{1})$, which is half of the
solid angle enclosed by the orange-slice-shaped path on the Bloch
sphere  and independent of the operation time. As a result, the
desired SGQG $U_1=\exp(i\gamma_1\sigma_z)$ is realized.

We then show how to achieve the gate $U_{2}$ geometrically in three
steps via an ``orange slice" scheme analogous to the gate $U_1$. In
Fig. 1(b), we plot the cyclic evolution path ($BCDAB$) on the Bloch
sphere surface in which point $B$ (corresponding to $\chi=\pi/2$) is
the initial state $|\lambda_{+}\rangle$. From $t=0$ to $t=\tau$, the
Rabi frequency and detuning of the original Hamiltonian
$H_{0}(0,\tau)$ with a constant phase $\varphi^{\prime}_{1}$ are
given by
\begin{equation}
\Omega_{R}(t)=\Omega_{0}[1+\cos(\frac{\pi
t}{\tau})],\Delta(t)=\Delta_{0}[\cos(\frac{\pi t}{\tau})-1].
\end{equation}
The system is driven to point $C$ by superadiabatic Hamiltonian
$H_{s}(0,\tau)$ with a constant phase $\varphi^{\prime}_{1}$.
Thereafter, from $t=\tau$ to $t=3\tau$, the parameters of the
Hamiltonian $H_{0}(\tau,3\tau)$ with a constant phase
$\varphi^{\prime}_{2}$ are given by
\begin{equation}
\Omega_{R}(t)=
\begin{cases}
\Omega_{0}[1-\cos(\frac{\pi (t-\tau)}{\tau})], &\mbox{$\tau\leq t<2\tau$}\\
\Omega_{0}[1+\cos(\frac{\pi (t-2\tau)}{\tau})], &\mbox{$2\tau\leq,
t\leq3\tau$}
\end{cases}
\end{equation}
and
\begin{equation}
\Delta(t)=
\begin{cases}
\Delta_{0}[\cos(\frac{\pi (t-\tau)}{\tau})+1], &\mbox{$\tau\leq t<2\tau$}\\
\Delta_{0}[\cos(\frac{\pi (t-2\tau)}{\tau})-1], &\mbox{$2\tau\leq,
t\leq3\tau$}.
\end{cases}
\end{equation}
The system is driven to point $A$ by superadiabatic Hamiltonian
$H_{s}(\tau,3\tau)$ with a constant phase $\varphi^{\prime}_{2}$.
Finally, from $t=3\tau$ to $t=4\tau$, the parameters in the original
Hamiltonian $H_{0}(3\tau,4\tau)$ with phase $\varphi^{\prime}_{1}$
are given by
\begin{eqnarray}
\begin{array}{ll}
\Omega_{R}(t)=\Omega_{0}[1-\cos(\frac{\pi (t-3\tau)}{\tau})],\\
\Delta(t)=\Delta_{0}[\cos(\frac{\pi (t-3\tau)}{\tau})+1].
\end{array}
\end{eqnarray}
The system is driven back to point $B$ by superadiabatic Hamiltonian
$H_{s}(3\tau,4\tau)$ with phase $\varphi^{\prime}_{1}$. In these
three steps, the $|\lambda_{+}\rangle$ state evolves along the paths
$BC$, $CA$ and $AB$ on the Bloch sphere and finally returns to the
starting point $B$ to form a cyclic loop. Similar to the proof in
$U_1$, we can show that the dynamical phases accumulated in these
three paths are completely canceled out. Meanwhile, the required
geometric phase
$\gamma_{2}=\pi-(\varphi^{\prime}_{2}-\varphi^{\prime}_{1})$. As a
result, the desired SGQG $U_2=\exp(i\gamma_2\sigma_x)$ is achieved.


We now turn to implementing a nontrivial two-qubit gate. We adopt a
system with the level structure similar to those used in the recent
dynamic \cite{Jelezko} and holonomic \cite{Zu} experiments, which
exploit the NV center electron spin as the target qubit and one
nearby $^{13}$C nuclear spin as the control qubit (with the
computational basis states $|\uparrow\rangle$ and
$|\downarrow\rangle$ ). The single electron spin is coupled to a
single $^{13}$C nuclear spin through hyperfine interaction, and the
resultant level configuration is shown in Fig. 1(c). The different
levels can be coupled by state-selective microwave and
radio-frequency fields. In particular, by applying the microwave
field $\mathrm{MW}_{tq}$ with adjustable frequency, Rabi frequency,
and phase, we can realize the superadiabatic coupling Hamiltonian
(\ref{Hs}) on the two-dimensional subspace spanned by
$\{|0,\uparrow\rangle,|1,\uparrow\rangle\}$ of the computational
space spanned by
$\{|0,\downarrow\rangle,|1,\downarrow\rangle,|0,\uparrow\rangle,|1,\uparrow\rangle\}$.
As with the single-qubit gate (\ref{SU}) above, we can achieve the
following superadiabatic geometric two-qubit gate:
\begin{equation}
\label{Utq} U_{tq}=|\uparrow\rangle\langle\uparrow|\otimes
U(\chi_{tq},\gamma_{tq})+|\downarrow\rangle\langle\downarrow|\otimes
I,\end{equation}
 where $I$ denotes the unit $2\times2$ matrix. The unwanted effect caused by the
 microwave coupling with the subspace of the nuclear spin pointing
 downward
 can be neglected when the hyperfine coupling $A$ is larger than
 $2\pi$ $\times$ 100 MHz under the chosen parameters below, as
 shown in the Supplemental Material \cite{Sup}.
 As a typical case, we can precisely realize the superadiabatic
geometric controlled-NOT gate  by controlling the microwave field
$\mathrm{MW}_{tq}$ just as with the implementation of $U_{2}$ (i.e.,
$\chi_{tq}=\pi/2$) and by choosing the geometric phase
$\gamma_{tq}=\pi/2$. Furthermore, the superadiabatic geometric
controlled-PHASE gate $U_{cp}$ can be achieved when
$U(\chi_{tq},\gamma_{tq})=U_1$, i.e., $\chi_{tq}=0$. Therefore,
$U_{tq}$ is sufficient for universal quantum computation when
assisted by a combination of the gates $U_{1}$ and $U_{2}$.

\begin{figure}[ptb]
\begin{center}
\includegraphics[width=10cm]{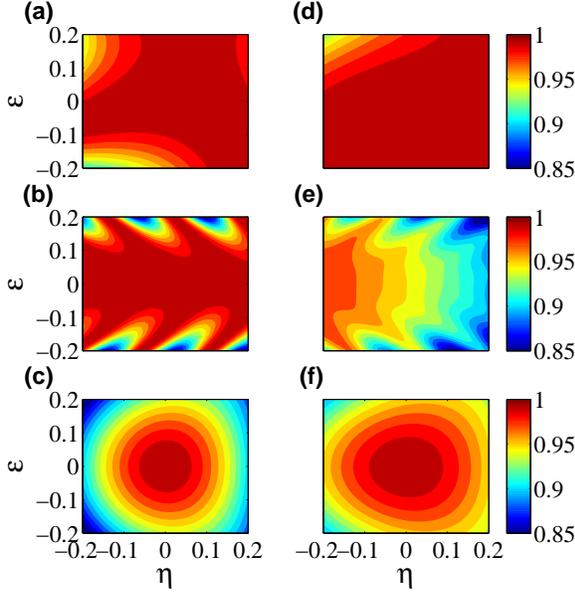}
\caption{\label{fig:charateristic}(color online). The fidelities $F$
versus the deviations of Rabi frequency $\eta\Omega_{sm}$ and
frequency detuning $\epsilon\Omega_{sm(nh)}$ from the ideal gates.
(a) Superadiabatic geometric $U_1$ with operation time $T_{sa}=0.64$
$\mu$s. (b) Adiabatic geometric $U_1$ with $T_{a}=10T_{sa}$. (c)
Nonadiabatic holonomic $U_1$ with $T_{nh}=T_{sa}$. (d)
Superadiabatic geometric $U_{cp}$ with operation time
$T^{\prime}_{sa}=0.64$ $\mu$s. (e) Adiabatic geometric $U_{cp}$ with
$T^{\prime}_{a}=10T^{\prime}_{sa}$. (f) Nonadiabatic holonomic
$U_{cp}$ with $T^{\prime}_{nh}=T^{\prime}_{sa}$.}
\end{center}
\end{figure}

We now discuss the performance of SGQGs as compared with  two other
kinds of geometric gates.  The imperfection of a quantum gate  is
usually due to the fluctuations of the control fields and the NV
center environments, specifically,  the fluctuations of the
amplitude, phase, and frequency of the control microwave fields as
well as the nearby nuclear spins ($^{13}$C and the host $^{15}$N)
and far-away random spin bath in the NV center environments. All of
these fluctuations may be classified as high- or low-frequency
noises based on the changes in the operation time of the quantum
gates. The random spin bath can be considered as a high-frequency
noise, and the nearby nuclear spins and the fluctuations of the
control fields can be approximately taken as quasistatic noises, a
low-frequency noise \cite{Rong}. Previous theories
\cite{Chiara,Zhu1} and experiments \cite{Berger} have shown that
geometric quantum gates are robust against high-frequency
perturbations, and in Supplemental Material \cite{Sup}, we confirm
that SGQGs keep this merit. Therefore, here we focus on a typical
low-frequency noise, the systematic errors in the control
parameters, which effectively include all quasistatic noises.

We choose SGQGs $U_{1}$ and $U_{cp}$ as the test cases, comparing
them with the adiabatic geometric phase gate without the
superadiabatic correction Hamiltonian $H_1$, and with the
nonadiabatic holonomic phase gate experimentally realized in Ref.
{\cite{Zu}}. As for $U_1$, the isolating two-level system as shown
in Fig. 1(a) is studied, with the parameters
$\Omega_{0}=2\pi\times2$ MHz and $\Delta_{0}=6$ MHz, which have been
used to realize the adiabatic geometric phase gate in Refs.
{\cite{Wu,Note}}. The maximum superadiabatic Rabi frequency
$\Omega_{sm}$ is a function of $\tau$. Moreover, in order to
guarantee that $\Omega_{sm}$ is not larger than the peak Rabi
frequency $\Omega_{Rm}$ of the original Hamiltonian $H_{0}$, we set
$\tau=0.16$ $\mu$s. The operation time $T_{sa}$ of the
superadiabatic geometric phase gate is $4\tau=0.64$ $\mu$s.
The nonadiabatic holonomic phase gate is realized by setting the
parameters $\theta=0$ and $\varphi=0$ of the three-level $\Lambda$
system Hamiltonian
$H=\hbar\Omega(t)[\sin(\theta/2)e^{i\varphi}|e\rangle\langle0|-\cos(\theta/2)|e\rangle\langle1|+\mathrm{H}.\mathrm{c}.]$,
as studied in Refs. \cite{Arroyo-Camejo,Zu,Abdumalikov,ES}. The
envelope $\Omega(t)$ is designed as a truncated Gaussian pulse
described by $\Omega_{nh}\exp(-t^{2}/\sigma^{2})$. We choose
$\Omega_{nh}=\Omega_{sm}/1.134$ as the peak Rabi frequency and
$2\sigma=4\sqrt\pi/\Omega_{nh}$, so the operation time $T_{nh}$ of
the nonadiabatic holonomic phase gate is the same as $T_{sa}$.

To implement the $U_{cp}$, we  control the microwave field
$\mathrm{MW}_{tq}$ coupled with the subspace
$\{|0,\uparrow\rangle,|1,\uparrow\rangle\}$ to change just as the
three different phase gates discussed above; however, the hyperfine
coupling between electron spin and $^{13}$C nuclear spin is chosen
as $2\pi$ $\times$ 127 MHz. In the numerical calculation of the
fidelities $F$ below, the effects of
$\{|0,\downarrow\rangle,|1,\downarrow\rangle\}$ are taken into
account.
All the above parameters are experimentally achievable on the NV
center of the recent experiments \cite{Jelezko}.

The quality of single-qubit (two-qubit) gates is characterized by
the ``intrinsic fidelity" (fidelity neglecting decoherence)
$F=|\mathrm{Tr}(UU_{0}^{\dag})|/2$
[$|\mathrm{Tr}(UU_{0}^{\dag})|/4$] \cite{Neilsen}, where $U$ and
$U_{0}$ are the operators for an imperfect and ideal single-qubit
(two-qubit) gate, respectively. Figure. 2 shows the simulated
fidelities of the three different kinds of phase (controlled-PHASE)
gates as a function of the relative Rabi frequency deviation
$\eta=\Delta\Omega_{sm}/\Omega_{sm}$, where $\Delta\Omega_{sm}$ is
the deviation from its center value $\Omega_{sm}$, and relative
frequency detuning $\epsilon=\delta/\Omega_{sm(nh)}$, where $\delta$
is the static frequency detuning. Clearly, the SGQG $U_1$ ($U_{cp}$)
is more robust against systematic errors than the nonadiabatic
holonomic $U_1$ ($U_{cp}$) as shown in Figs. 2(a) and 2(c) [Figs.
2(d) and 2(f)]. Remarkably, the adiabatic geometric $U_1$ ($U_{cp}$)
without superadiabatic correction Hamiltonian $H_{1}$ is less robust
than the SGQG even when the operation time
$T^{(\prime)}_{a}=10T^{(\prime)}_{sa}$, as shown in Figs. 2(a) and
2(b) [Figs. 2(d) and 2(e)] \cite{Note}. Rather than at $\eta=0$, the
highest fidelity in Fig. 2(e) is at the negative $\eta$, since the
effect from $\{|0,\downarrow\rangle,|1,\downarrow\rangle\}$ is
larger when $\Omega_{sm}$ is larger. Therefore, our SGQG has both
fast and robust features that are significant in quantum
manipulation.



In conclusion, we have proposed a general scheme to realize
universal SGQGs, with application to the NV center system as an
example. The designed universal gates are based on geometric phases,
which can be robust against certain stochastic errors, such as
fluctuations of the driving fields in realistic situations
\cite{Berger,Chiara,Zhu1}. The evolutions are superadiabatic and can
be fast and robust against the systematic errors. The physical
implementation of the scheme can be realized in the NV center
systems with current technology.
Therefore, it is promising to experimentally implement these
robust and fast universal SGQGs on NV center qubits at room
temperature.

\begin{acknowledgments}

This work was supported by the NSF of China (Grants No. 11474107,
No. 11125417,  and No. 11474153),   the GNSFDYS (Grant No.
2014A030306012), the FOYTHEG (Grant No. Yq2013050), the PRNPGZ
(Grant No. 2014010), and the PCSIRT (Grant No. IRT1243). L.Z.T. was
also supported by the SRFGS of SCNU.

\end{acknowledgments}


\begin{thebibliography}{99}
\bibitem{Note} The adiabatic condition is usually well satisfied under the condition $\Omega_{max}\times T>2\pi\times10$ with $T$ the operation time,
which is comparable with $\Omega_{max}\times T_a=2\pi\times25.6$ for
the parameters $\Omega_{max}=2\Omega_{0}=2\pi \times 4$ MHz and
$T_a=6.4 \mu$s here. So the fidelity of the adiabatic geometric
phase gate is pretty high under such parameters. If the operation
time decreases and thus the adiabatic condition is not satisfied,
the fidelity of the adiabatic geometric phase gate will decrease
quickly even in the absence of any noise.
\end{thebibliography}
\end{document}